# Effect of Mg substitution in $Sr_2SiO_4:Eu^{2+}$ nanophosphors for blue and white emission at near UV excitation

**Sumit Dubey[a], Pratik Deshmukh[a], S. Satapathy[a, *], M. K. Singh and P. K. Gupta[a]**

[a] Nano Functional Materials Laboratory, Laser Materials Development & Devices Division, Raja Ramanna Centre for Advanced Technology, Indore 452013, India

**Corresponding author**

[*] **E- mail:** srinu73@cat.ernet.in, srinusatapathy@gmail.com

Phone: 91 731 2488660/8658

Fax: 91 731 2488650

## Abstract

Nanophosphors of $(Sr_{0.98-x}Mg_xEu_{0.02})_2SiO_4$ (x=0, 0.18, 0.38, 0.58 and 0.78) were prepared through low temperature solution combustion method and their luminescence properties were studied. The emission peak for $Eu^{2+}$ doped $Sr_2SiO_4$ nanophosphor is observed at ~490 nm and ~553 nm corresponding to two $Sr^{2+}$ sites Sr(I) and Sr(II) respectively for 395 nm excitation but the addition of $Mg^{2+}$ dopant in $Sr_2SiO_4$ leads to suppression of ~553 nm emission peak due to absence of energy levels of Sr (II) sites which results in a single broad emission at ~460nm. It was shown that the emission peak blue shifted with increase in Mg concentration which may be attributed to change in crystal field environment around Sr(I) sites. Therefore the $(Mg_{0.78}Sr_{0.20}Eu_{0.02})_2SiO_4$: $Eu^{2+}$ nanophosphor can be used for blue emission and the $Sr_2SiO_4:Eu_{0.04}^{2+}$ for green-yellow emission at 395 nm excitations. The CIE coordinates for mixed powders of $(Mg_{0.78}Sr_{0.20}Eu_{0.02})_2SiO_4$ and $Sr_2SiO_4:Eu_{0.04}^{2+}$ (in 1: 1 ratio) fall in the white region demonstrating the possible use of the mixture in white light generation using near UV excitation source.

## Introduction

Nanophosphor typically solid inorganic materials have become popular from the last decade due to their extraordinary and unique physical, chemical and optical properties. Nanophosphors are also preferable in white LED applications due to lower scattering loss as compared to larger particle sized phosphors. [1] Phosphor based white LEDs is an efficient source of light because of its excellent properties such as high brightness, low power consumption, fast response time and long life time. [2-3] One can't realize a phosphor based white LED without blue excitation source or blue emission phosphor. [4] The utilization of near ultraviolet (UV) radiation as excitation source for blue emitting phosphor is beneficial as human eye is not very sensitive to the wavelengths range ~380 to ~ 400 nm [5] and also, the utilization of near-UV excitation source for blue emission will provide greater efficiency because stokes shift is small; hence lesser energy is converted into heat. Therefore, use of phosphors that can be excited in near UV region is the basis of research nowadays.

Rare earth doped Strontium Silicate based hosts ($Sr_2SiO_4$) have been studied in details to provide greater efficiency compared to the conventional YAG:Ce based white LED. [6-8] Divalent Europium($Eu^{2+}$) is a popular rare-earth element which on doping in hosts like phosphates, silicates, aluminates, shows parity-allowed 4f-5d energy level transition from ultraviolet to red based on host lattice and co-valence in the host.[9-10] The $Eu^{2+}$ doped $Sr_2SiO_4$ is widely being used for producing yellow light using blue or UV LEDs excitation source.[6] The $Sr^{2+}$ in $Sr_2SiO_4$ have two cation sites Sr(I) and Sr(II) which are ten and nine coordinated by oxygen atoms respectively. Thus doping of $Eu^{2+}$ ions in $Sr_2SiO_4$, results in two emission bands at ~490 nm and ~560 nm. These two transitions are attributed to $^4f_65d \rightarrow {}^4f_7$ transition of $Eu^{2+}$. [11-12] The preference of $Eu^{2+}$ for Sr(I) or Sr(II) in $Sr_2SiO_4$ at different excitation wavelengths was also reported in the literature.[13] The effects of $Mg^{2+}$ doping in $Sr_2SiO_4:Eu^{2+}$ was studied at 365 nm excitation and its influence on the two emission bands i.e. ~ 459 nm and ~564 nm has also been reported. [14]

The silicate based blue phosphor $(Sr_xMg_{1-x}\ Eu_{0.02})_2Si_{1.02}O_{4.08}$ prepared by solid state route for 400 nm excitation, showed two emission bands at ~460 nm and ~570 nm for x = 0.6 and above Sr doping concentrations for 400 nm excitation.[4] However, no studies have yet examined the photoluminescence characteristics of nano sized $(Sr_{0.98-x}Mg_xEu_{0.02})_2SiO_4$ phosphor for near UV excitation. Thus, this work is carried out to investigate the effects on photoluminescence (PL) characteristics of $(Sr_{0.98-x}Mg_xEu_{0.02})_2SiO_4$ nanophosphor by varying Mg doping concentrations.

## Experimental

Nano powders of $(Sr_{0.98-x}Mg_xEu_{0.02})_2SiO_4$ (x=0, 0.18, 0.38, 0.58 and 0.78) were synthesized by combustion method. Nitrate solutions, $Sr(NO_3)_2$, $Mg(NO_3)_2$ and $Eu(NO_3)_3$ were prepared based on stoichiometric ratios using carbonates of Sr and Mg, and $Eu_2O_3$ respectively. Tetraethoxysilane (TEOS) solution was mixed with nitrate solution. The solution was stirred for 2 hours and Glycine was added into the solution which acts as fuel in the combustion. The solution was heated at 500°C for combustion. The powder obtained from combustion was calcined at 900°C in a reducing atmosphere (90% $N_2$+ 10% $H_2$) for two hours to get desired phase.

The confirmation of phase of the synthesized phosphors was done using Rigaku X-Ray diffractometer (Powder-XRD) with Cu K$\alpha_1$ ($\lambda$= 0.15405 nm) radiation. The particle size and morphology of phosphors were observed by field emission scanning electron microscopy (FESEM; ZEISS-SIGMA). The PL and photoluminescence-excitation (PLE) spectra were measured using a FLS920-S steady state Fluorometer, Edinburgh Instruments Ltd. The standard light source used for excitation is a 450W ozone free xenon arc lamp.

## Results

The XRD Patterns of the as-prepared $(Sr_{0.98-x}Mg_xEu_{0.02})_2SiO_4$ phosphors were shown in **figure 1a**. The diffraction peaks of the prepared samples were indexed using JCPDS file **(39-1256)** which confirms the proper phase formation of $Sr_2SiO_4$:$Eu_{0.04}^{2+}$. The pure $Sr_2SiO_4$:$Eu^{2+}$ belongs to the orthorhombic phase of α-$Sr_2SiO_4$. The replacement of Sr by Mg is also reflected in XRD pattern. The $(Sr_{0.98-x}Mg_xEu_{0.02})_2SiO_4$ remains in orthorhombic phase with small percentage of β-$Sr_2SiO_4$ upto x = 0.58, [confirmed from the existence of the peak at 32.8$^o$ according to JCPDS file 38-0271 (marked with circle in the XRD pattern of **figure 1a**)]. Furthermore, we observed shifting of diffraction peaks towards higher 2θ value due to doping of Mg (as shown in **figure 1b**). This may be due to the substitution of $Sr^{2+}$ ions by smaller $Mg^{2+}$ ions, which results in shrinkage of lattice parameter of $Sr_2SiO_4$. The XRD pattern of $(Sr_{0.20}Mg_{0.78}Eu_{0.02})_2SiO_4$ shows more extra peaks in addition to XRD peaks of $Sr_2SiO_4$. These peaks are assigned to (021), (101), (111) and (002) planes of orthorhombic $Mg_2SiO_4$ (JCPDS: **78-1371**). This confirms separate phase formation of $Mg_2SiO_4$ along with $Sr_2SiO_4$ orthorhombic phase for x = 0.78. Moreover we found one extra small peak at **28.2°** in all diffraction patterns which may be due to unconverted phase of $Eu^{3+}$.[15]

The particle size and morphology of the prepared powders were analyzed using FE-SEM. The particles appear to be nearly spherical for $Sr_2SiO_4$ and no significant change has been observed in its morphology with subsequent doping of Mg in it (**figure 2**). The particles sizes were found to be in the range of ~20-80 nm. The deviation from spherical shape is due to nature of synthesis i.e. combustion method, where control over particle size and shape is very difficult. The inset figures are enlarged images of small nano particles. Some large particles along with small nano particles were observed in SEM images and close observation reveals that these large particles are agglomerated blocks of small particles.

The PLE spectra of $(Sr_{0.98-x}Mg_xEu_{0.02})_2SiO_4$ phosphor are shown in **figure 3** which were obtained by placing detector at 460 nm. The PLE spectra confirm that the sample can be excited using broadband excitation from 250 to 400 nm**.** A broad peak at 313nm is observed in PLE spectrum of $Sr_2SiO_4{:}Eu_{0.04}^{2+}$ **(figure 3)** and the single broad peak splits into two broad peaks at ~280 nm and ~340nm due to doping of $Mg^{2+}$. Therefore we have investigated luminescence properties of the phosphors at 280 and 340nm excitation. We also studied the PL properties of these nanophosphors at 395 nm. Although the 395 nm does not correspond to typical excitation peak in PLE spectra but near UV LED of this wavelength is easily available.

The PL spectrum of $Sr_2SiO_4{:}Eu_{0.04}^{2+}$ excited at 395 nm is shown in **figure 4** which shows an asymmetric broad spectrum with two broad peaks at ~480 and ~553 nm. The peak at ~553 nm is found to be more intense than the peak at ~480 nm The PL spectra for $(Sr_{0.98-x}Mg_xEu_{0.02})_2SiO_4$ (x= 0.18, 0.38, 0.58 and 0.78) show a symmetric broad peak at ~460 nm **(figure 4).** It is also observed that the emission peak is blue shifted with increase in concentration of $Mg^{2+}$ dopant. Maximum PL intensity observed for $(Sr_{0.20}Mg_{0.78}Eu_{0.02})_2SiO_4$ nanophosphor

The PL spectra for all compositions for 340 nm excitation are shown in **figure 5**. The emission peaks were blue shifted with subsequent increase in Mg dopant concentrations in $(Sr_{0.98-x}Mg_xEu_{0.02})_2SiO_4$ (x= 0.18, 0.38, 0.58 and 0.78). In case of $Sr_2SiO_4:Eu_{0.04}^{2+}$, an asymmetric broad emission spectrum was observed. The emission peak at ~490 nm for $Sr_2SiO_4:Eu_{0.04}^{2+}$ dominated in the broad spectrum but the emission peak at ~550 nm was suppressed **(figure 5).** The smaller intensity peaks at ~575 and ~622 nm were also observed which may be due to very small unconverted $Eu^{3+}$ presence in the calcined sample.

The emission spectra of all compositions excited at 280 nm are shown in **figure 6.** The emission range was reduced i. e. from 375 to 540 nm in order to avoid second order diffraction peaks. The $Sr_2SiO_4:Eu_{0.04}^{2}$ shows broad emission centered at 489nm while peaks at 459, 461, 466 and 471 nm were observed for $(Sr_{0.98-x}Mg_xEu_{0.02})_2SiO_4$ corresponding to x=0.78, 0.58, 0.38 and 0.18 respectively for 280 nm excitation.

## Discussion

The excitation spectrum of $Sr_2SiO_4:Eu^{2+}$ shows a broad absorption band centered at 313 nm when PL monitored at 460 nm. In $Sr_2SiO_4:Eu^{2+}$ the $4f^65d^1$ electron state strongly interacts with neighbouring anion ligands and forms a broad absorption band called charge transfer bands (CTBs). So broad excitation band is observed due to the electronic transition to these CTBs of $Eu^{2+}$ ions ($4f^7 \rightarrow 4f^65d^1$).[16] Since broad excitation spectrum has been observed (from 250nm to 400nm) the phosphor can be excited by near UV region (i. e. 370-400nm). The excitation spectrum of Mg doped $Sr_2SiO_4:Eu^{2+}$ shows splitting of broad peak into two peaks due to addition of Mg.  These splitting of broad peak may be ascribed to CTBs associated with Sr(I) and Sr(II) sites in $Sr_2SiO_4$.

The broad PL emission centred at ~490 nm and at ~553 nm for $Sr_2SiO_4:Eu_{0.04}^{2+}$ under the excitation of 395 nm, is ascribed to the electric dipole-allowed transition of the $Eu^{2+}$ ions from the lowest level of the 5d excited state (CTBs at two different $Sr^{2+}$ sites) to the 4f ground state. The shorter wavelength emission band (~495nm) and the longer wavelength emission band (~553 nm) correspond to Sr(I) and Sr(II) site respectively. It was observed that the broad band at ~553 nm associated with Sr(II) disappears due to increase in Mg doping [i.e. in $(Sr_{0.98-x}Mg_xEu_{0.02})_2SiO_4$ (x=0.18, 0.38, 0.58 and 0.78) with increase in x) and the blue shifting of broad band at ~476 nm was also observed with increase in concentration of Mg in $Sr_2SiO_4$.

The ionic radius of $Eu^{2+}$ (~0.131 nm) is very similar to $Sr^{2+}$ (~0.132 nm) and much larger than $Si^{4+}$ (0.054 nm), makes way for $Eu^{2+}$ ions to substitute $Sr^{2+}$ ions in the crystal lattice.[17] The covalence of the Sr–O bonds in the lattice gets decreased when a part of $Sr^{2+}$ ion is substituted by $Mg^{2+}$ ions. So, the covalence of Eu–O bonds doped at the $Sr^{2+}$ (I) site gets decreased. This causes the difference between the 4f and 5d energy level to become larger since less negative charge transfers to the $Eu^{2+}$ ion. This makes the emission energy increased, which results in a blue-shift of the emission spectra. [9-10, 12]

According to the crystal field theory [9-10] in the host lattice of $Sr_2SiO_4$, the size of Sr(II) site is larger than that of Sr(I) Substitution of $Sr^{2+}$ ions by smaller $Mg^{2+}$ ion decreases the bond length. [14,18-19] The $Mg^{2+}$ ion may prefer for the Sr(I) site because the $Mg^{2+}$ ion is much smaller than $Sr^{2+}$. Hence, the crystal field strength could be improved more for Sr(I) when $Mg^{2+}$ ions are doped. Similarly, the symmetry of ligand ions surrounding $Eu^{2+}$ in the Sr(II) site maybe distorted due to large site size. Thus the crystal field effect cannot be efficiently exerted for this Sr(II) site which results in suppression of emission band at ~553 nm (related to Sr (II) sites).

Although, all the emission peaks were similar in case of Mg doped $Sr_2SiO_4:Eu_{0.04}^{2+}$ for 280, 340 nm and 395nm excitation but for pure $Sr_2SiO_4:Eu_{0.04}^{2+}$ the emission band is different for 395 nm excitation compared to excitation at 280 and 340 nm. An asymmetric broad emission band centred at ~490 nm has been observed for pure $Sr_2SiO_4:Eu_{0.04}^{2+}$ since the intensity of ~553nm peak decreases substantially for 280nm and 340nm excitation. This shows that for shorter wavelength UV excitation the luminescence due to Sr(I) site is dominant than Sr(II) site.[11, 13]

The variation of the Commission International de L'Eclairage (CIE) chromaticity coordinates of the phosphors with different Sr:Mg ratio are calculated based on the corresponding PL spectrum upon 395 nm excitation, and the results are summarized in **figure 7** and **Table 1**. The CIE coordinates for $Sr_2SiO_4:Eu_{0.04}^{2+}$ were found to be in greenish-yellow region and with increase in Mg:Sr ratio the graph shifted from green yellow to deep blue. Similarly the CIE diagram for 340 nm excitation was studied, and the results are summarized in **figure 8** and **Table 2**. It was found that with increase in Mg:Sr ratio the graph shifted from green towards deep blue. The white light for 395 nm excitation is obtained by mixing equal amount of blue ($Sr_2SiO_4:Eu_{0.04}^{2+}$) and yellow-green emitting phosphor $(Mg_{0.78}Sr_{0.20}Eu_{0.02})_2SiO_4$ **(figure 9)**. Therefore these phosphors can be used in generation of white light using commercially available 395 nm excitation sources.

## Conclusions

In the present work, $(Sr_{1-x}Mg_xEu_{0.02})_2SiO_4$ (x=0, 0.18, 0.38, 0.58 and 0.78) nanophosphors were prepared by combustion method. The formation of orthorhombic α-$Sr_2SiO_4$ phase was confirmed for $Sr_2SiO_4:Eu^{2+}$ nanophosphor; while orthorhombic α-$Sr_2SiO_4$ along with few percentage of monoclinic β-$Sr_2SiO_4$ was obtained for $Mg^{2+}$ doped $Sr_2SiO_4:Eu_{0.04}^{2+}$. FE-SEM analysis showed particles to be nearly spherical for $Sr_2SiO_4$ and no significant change in its morphology was observed with subsequent doping of Mg. In $Sr_2SiO_4$ the $Eu^{2+}$ dopant occupies two different $Sr^{2+}$sites (Sr (I) and Sr (II)) which will lead to two emission bands i.e. at ~490 and ~553 nm for 395 nm excitation respectively. However, an asymmetric broad emission band centred at ~490 nm could be seen for $Sr_2SiO_4:Eu_{0.04}^{2+}$ at 280 and 340 nm excitation. This confirms that the luminescence from Sr(I) site is dominant than Sr(II) site for shorter wavelength excitation. Addition of $Mg^{2+}$ ions in $Sr_2SiO_4:Eu^{2+}$ leads to shifting of emission peak to shorter wavelength ~460 nm (blue region) for at 395nm excitation which attributed to preference of Mg for Sr(I) site. Therfore Mg doped $Sr_2SiO_4:Eu_{0.04}^{2+}$ nanophosphor can be used as blue emitting phosphor for near-UV excitation. The CIE coordinates for mixture of blue $(Mg_{0.78}Sr_{0.20}Eu_{0.02})_2SiO_4$ and yellow-green ($Sr_2SiO_4:Eu_{0.04}^{2+}$) with 1: 1 ratio lies in white region for 395 nm excitation which indicates great potential of these nano phosphors for white light generation.

**Acknowledgement**

We are thankful to Mr. Ashish Kumar Singh, NFM Laboratory, LMDDD, RRCAT, Indore, for his help in synthesis of phosphors. Authors also would like to thank Mr. Gopal Mohod for his help in infrastructure development for experimental setups.

## Tables:

**Table 1. CIE chromaticity coordinates (x,y) for $(Sr_{0.98-x}Mg_xEu_{0.02})_2SiO_4$ phosphors excited at 395nm**

| Sample no. | Sample Composition | CIE Coordinates(x,y) |
|---|---|---|
| 1 | $(Mg_{0.78}Sr_{0.20})_2SiO_4:Eu_{0.04}^{2+}$ | (0.1402,0.0714) |
| 2 | $(Mg_{0.58}Sr_{0.40})_2SiO_4:Eu_{0.04}^{2+}$ | (0.1395,0.0780) |
| 3 | $(Mg_{0.38}Sr_{0.60})_2SiO_4:Eu_{0.04}^{2+}$ | (0.1371,0.1185) |
| 4 | $(Mg_{0.18}Sr_{0.80})_2SiO_4:Eu_{0.04}^{2+}$ | (0.1333,0.1725) |
| 5 | $Sr_2SiO_4:Eu_{0.04}^{2+}$ | (0.3644,0.4783) |

**Table 2. CIE chromaticity coordinates (x, y) for $(Sr_{0.98-x}Mg_xEu_{0.02})_2SiO_4$ phosphors excited at 340nm**

| Sample no. | Sample Composition | CIE Coordinates(x,y) |
|---|---|---|
| 1 | $(Mg_{0.78}Sr_{0.20})_2SiO_4:Eu_{0.04}^{2+}$ | (0.1435,0.0668) |
| 2 | $(Mg_{0.58}Sr_{0.40})_2SiO_4:Eu_{0.04}^{2+}$ | (0.1384,0.0923) |
| 3 | $(Mg_{0.38}Sr_{0.60})_2SiO_4:Eu_{0.04}^{2+}$ | (0.1425,0.1057) |
| 4 | $(Mg_{0.18}Sr_{0.80})_2SiO_4:Eu_{0.04}^{2+}$ | (0.1412,0.1595) |
| 5 | $Sr_2SiO_4:Eu_{0.04}^{2+}$ | (0.1914,0.3021) |

**Figure captions:**

**Figure 1.** (a) XRD patterns of $(Sr_{0.98-x}Mg_xEu_{0.02})_2SiO_4$ phosphors (0≤ x≤ 0.78); (b) Magnified XRD pattern of $(Sr_{0.98-x}Mg_xEu_{0.02})_2SiO_4$ phosphors ($0 \le x \le 0.78$);

**Figure 2.** FE-SEM images of $(Sr_{0.98-x}Mg_xEu_{0.02})_2SiO_4$ phosphors [(A) x=0.78, (B) x=0.58, (C) x=0.38, (D) x=0.18, (E) x=0]. Inset figures show zoomed images of nano particles.

**Figure 3.** PLE spectra $(Sr_{0.98-x}Mg_xEu_{0.02})_2SiO_4$ phosphors. The inset shows the PLE spectra of $(Sr_{0.80}Mg_{0.18}Eu_{0.02})_2SiO_4$ nanophosphor.

**Figure 4.** PL spectra of $(Sr_{0.98-x}Mg_xEu_{0.02})_2SiO_4$ phosphors for 395nm excitation. The inset shows the PL spectra of $(Sr_{0.80}Mg_{0.18}Eu_{0.02})_2SiO_4$ phosphor.

**Figure 5.** PL spectra of $(Sr_{0.98-x}Mg_xEu_{0.02})_2SiO_4$ phosphors for 340nm excitation.

**Figure 6.** PL spectra of $(Sr_{0.98-x}Mg_xEu_{0.02})_2SiO_4$ phosphors for 280nm excitation. The inset shows the PL spectra of $(Sr_{0.80}Mg_{0.18}Eu_{0.02})_2SiO_4$ phosphor.

**Figure 7.** CIE chromaticity diagram of prepared $(Sr_{0.98-x}Mg_xEu_{0.02})_2SiO_4$ phosphors for 395nm excitation (A= $(Mg_{0.78}Sr_{0.20}Eu_{0.02})_2SiO_4$, B= $(Mg_{0.58}Sr_{0.40}Eu_{0.02})_2SiO_4$, C= $(Mg_{0.38}Sr_{0.60}Eu_{0.02})_2SiO_4$, D= $(Mg_{0.18}Sr_{0.80}Eu_{0.02})_2SiO_4$ , E= $Sr_2SiO_4:Eu_{0.04}^{2+}$ ).

**Figure 8.** CIE chromaticity diagram of prepared $(Sr_{0.98-x}Mg_xEu_{0.02})_2SiO_4$ phosphors for 340nm excitation (A= $(Mg_{0.78}Sr_{0.20}Eu_{0.02})_2SiO_4$, B= $(Mg_{0.58}Sr_{0.40}Eu_{0.02})_2SiO_4$, C= $(Mg_{0.38}Sr_{0.60}Eu_{0.02})_2SiO_4$, D= $(Mg_{0.18}Sr_{0.80}Eu_{0.02})_2SiO_4$ , E= $Sr_2SiO_4:Eu_{0.04}^{2+}$ )

**Figure 9.** CIE chromaticity diagram for combined $Mg_{0.78}Sr_{0.20}Eu_{0.02})_2SiO_4$ and $Sr_2SiO_4:Eu_{0.04}^{2+}$ nano phosphors indicating white point for 395nm excitation.

**Figure-1**

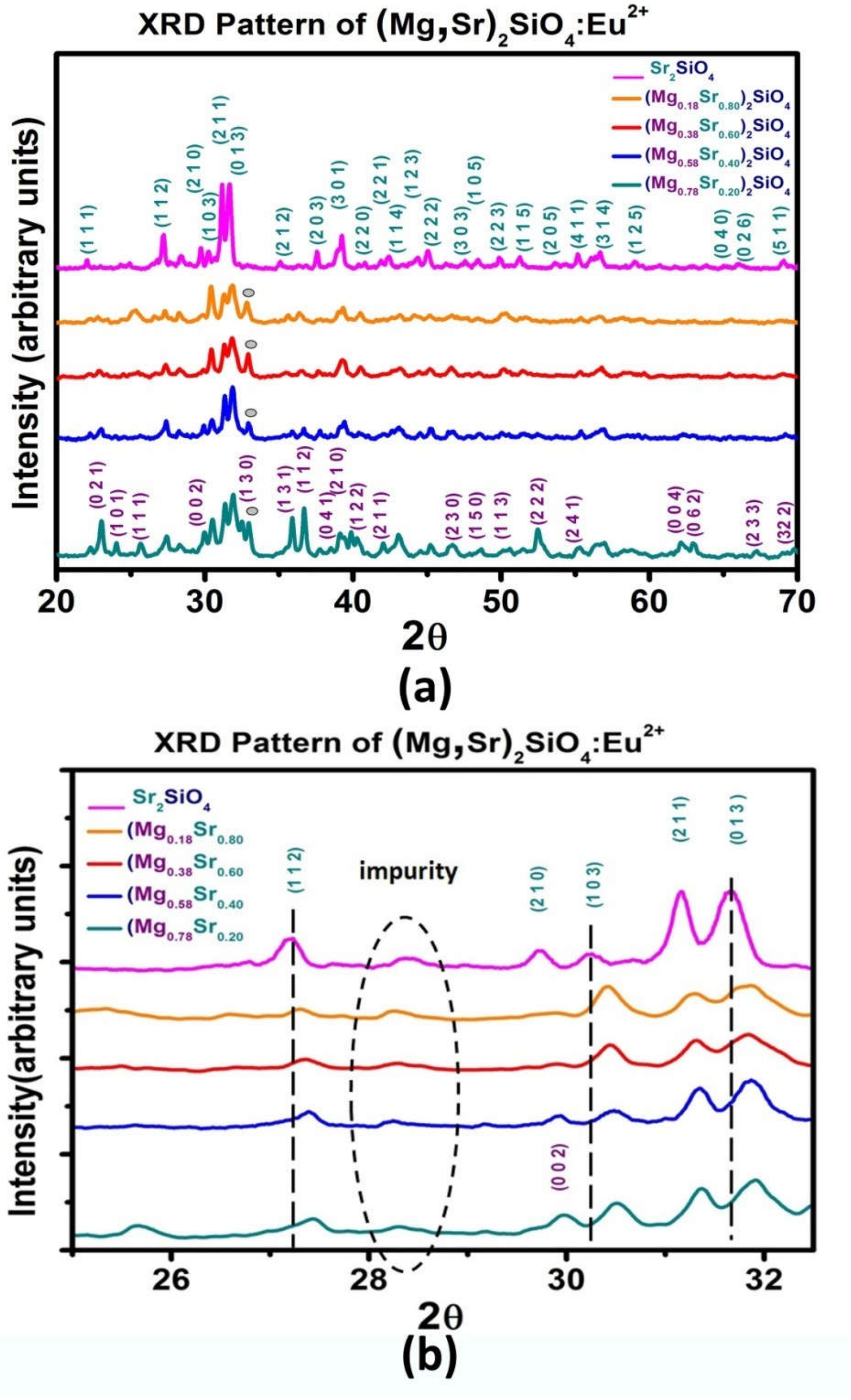

**Figure -2**

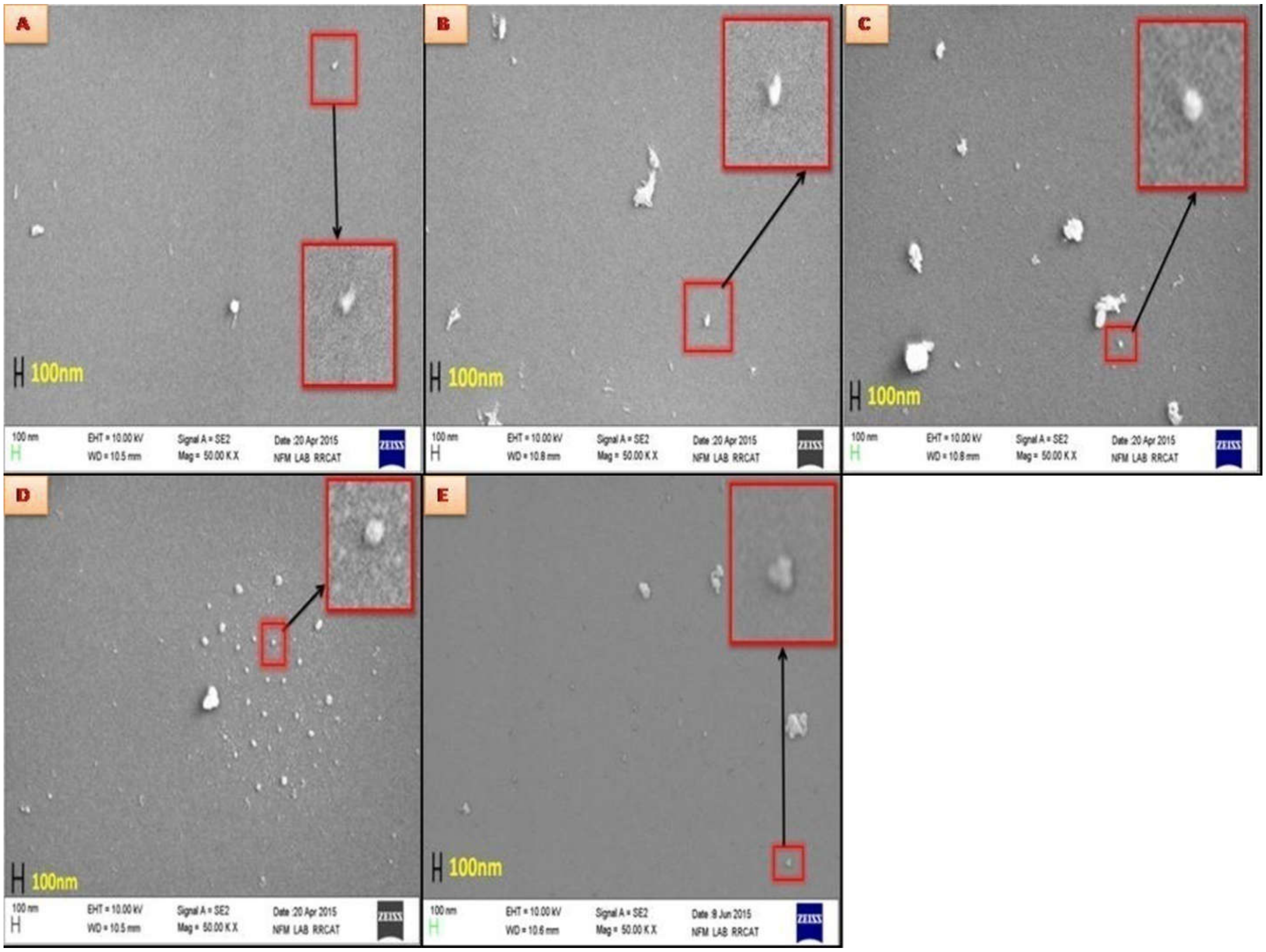

**Figure -3**

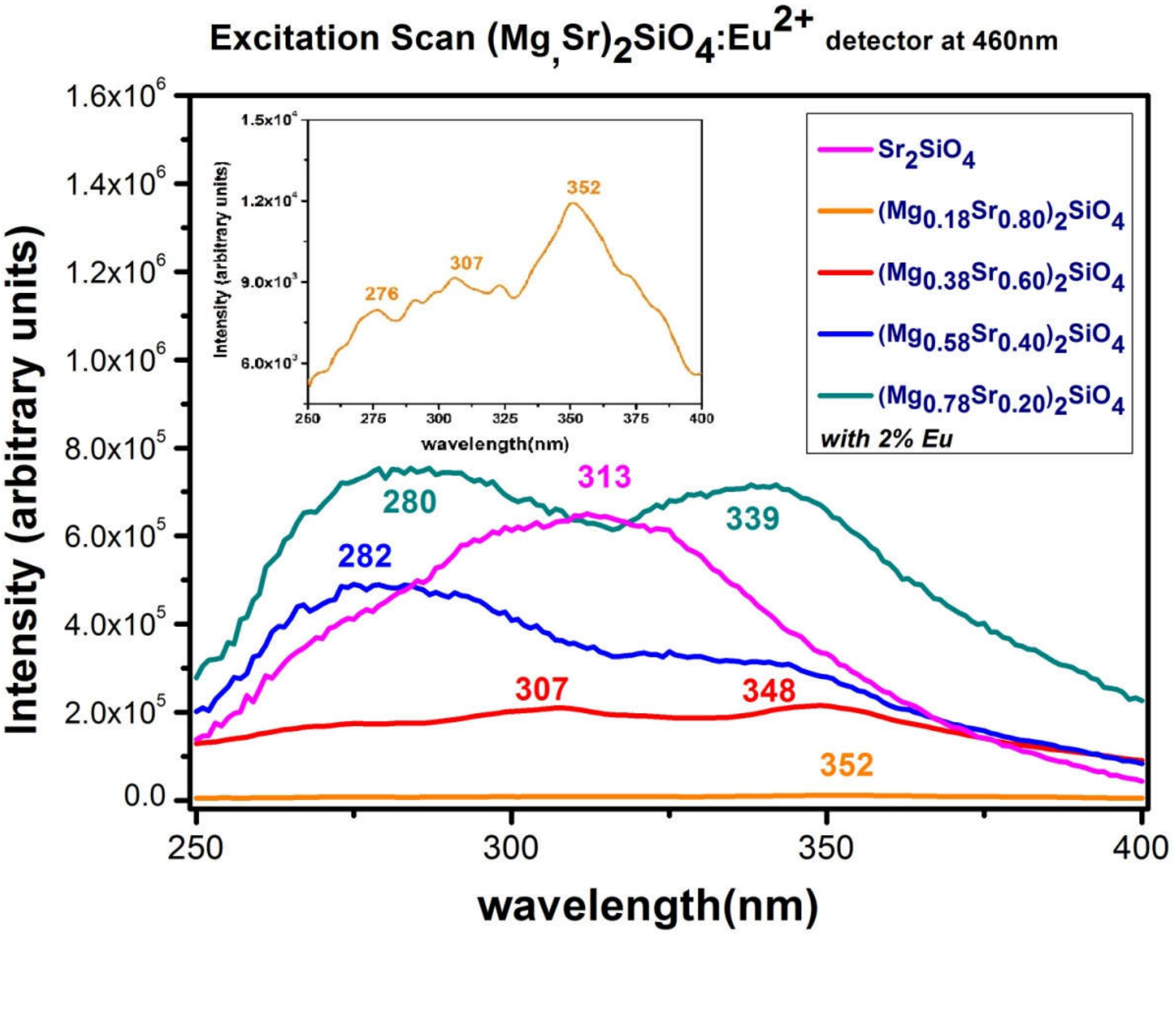

**Figure-4**

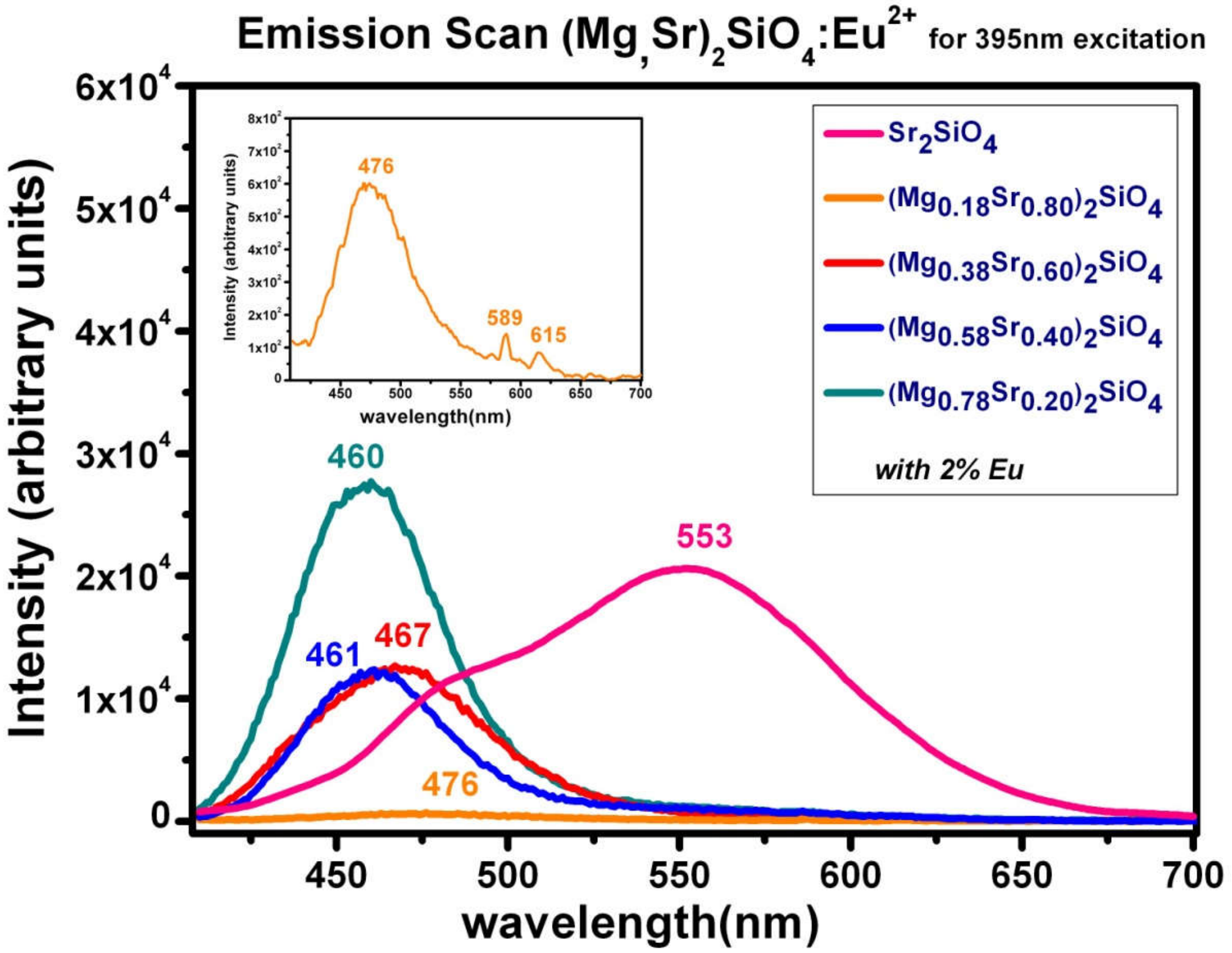

**Figure-5**

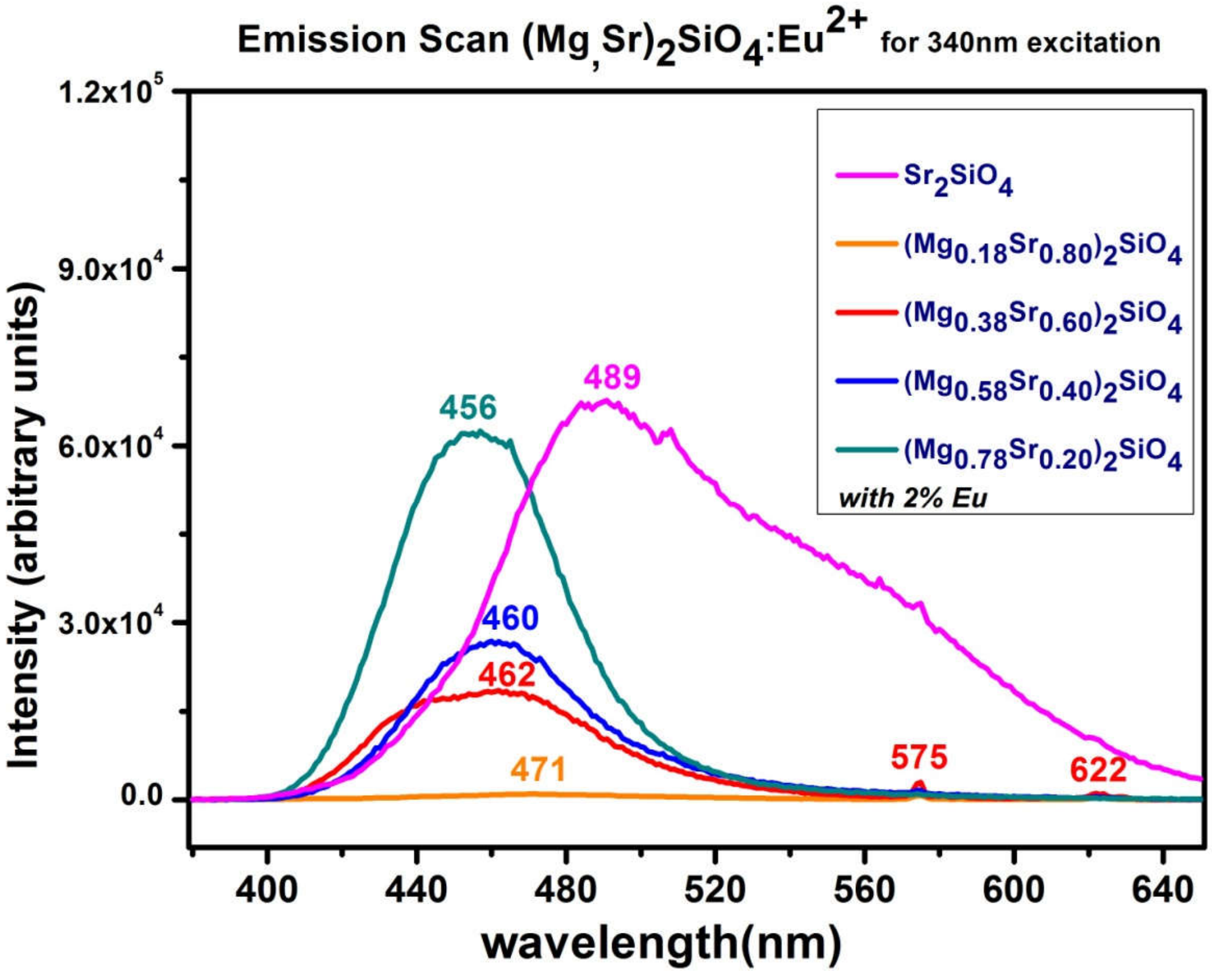

**Figure-6**

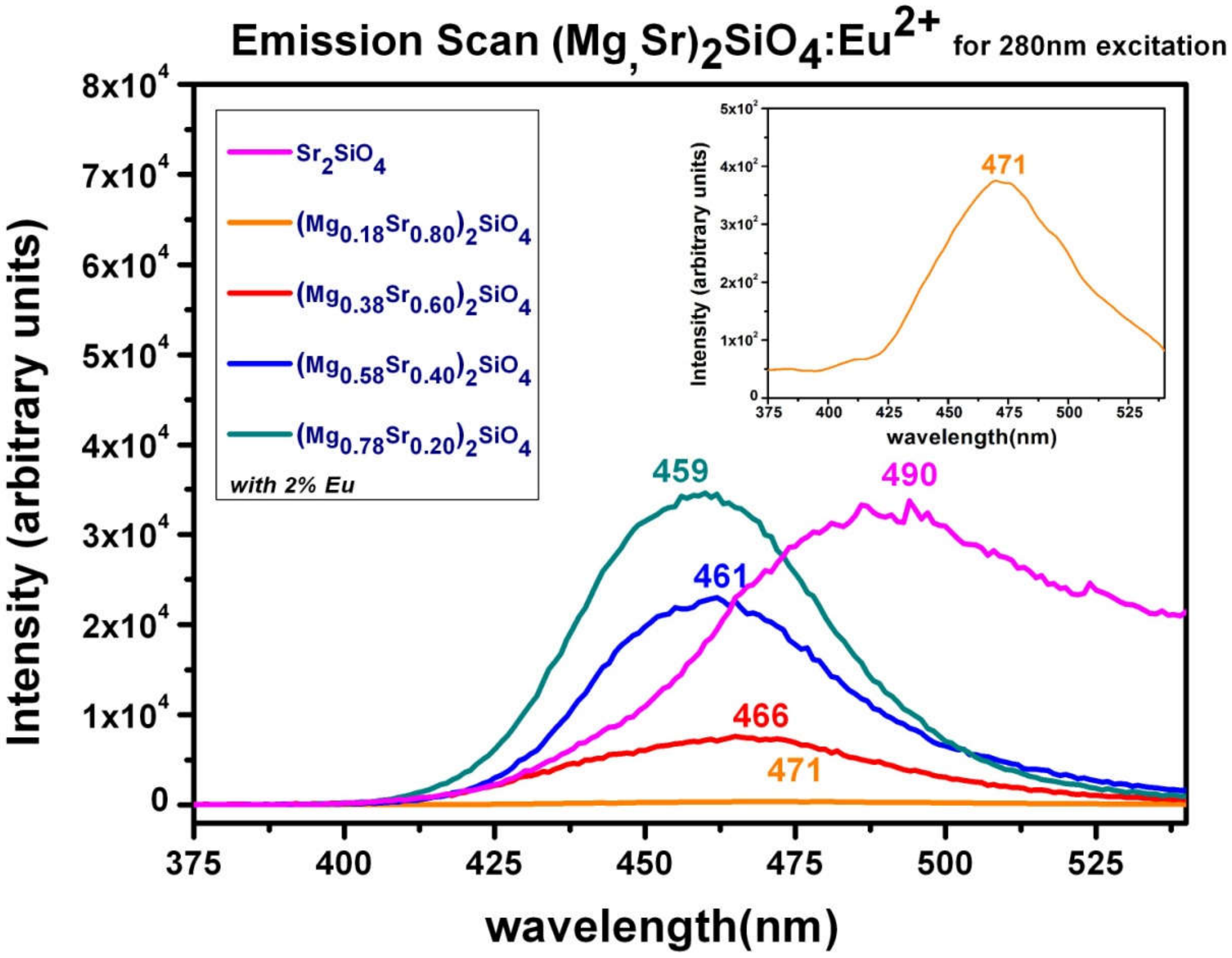

**Figure-7**

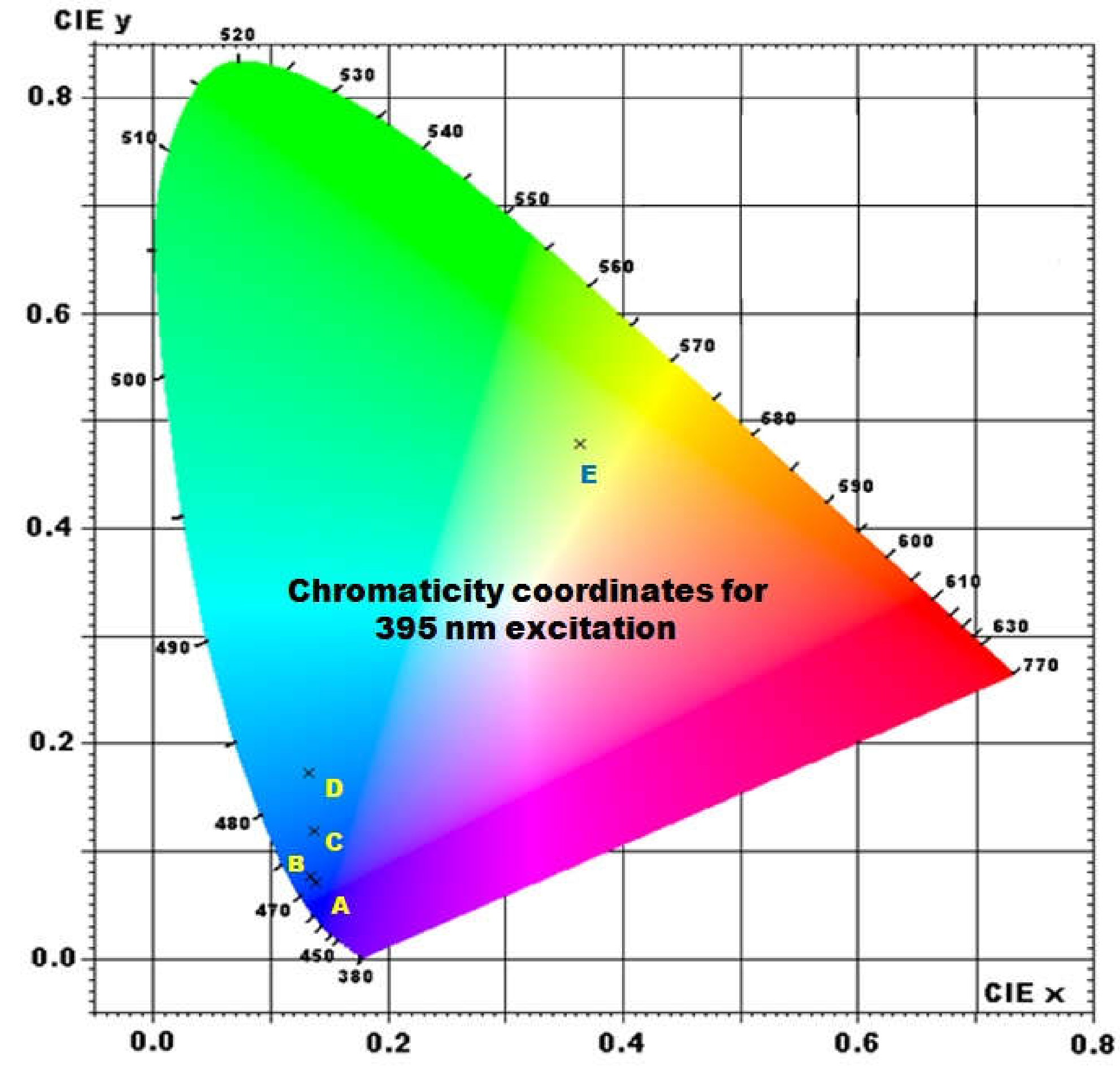

**Figure-8**

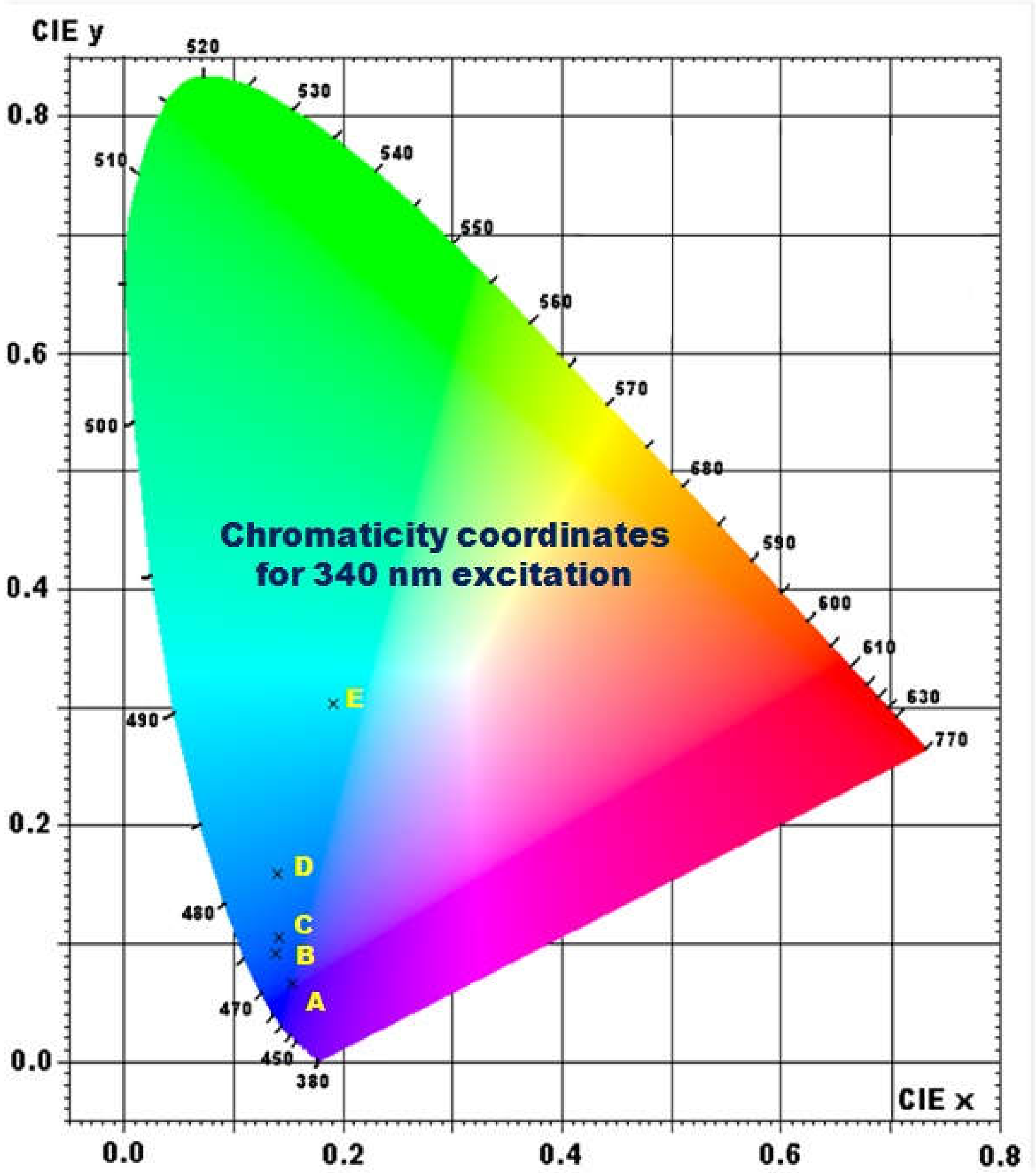

**Figure-9**

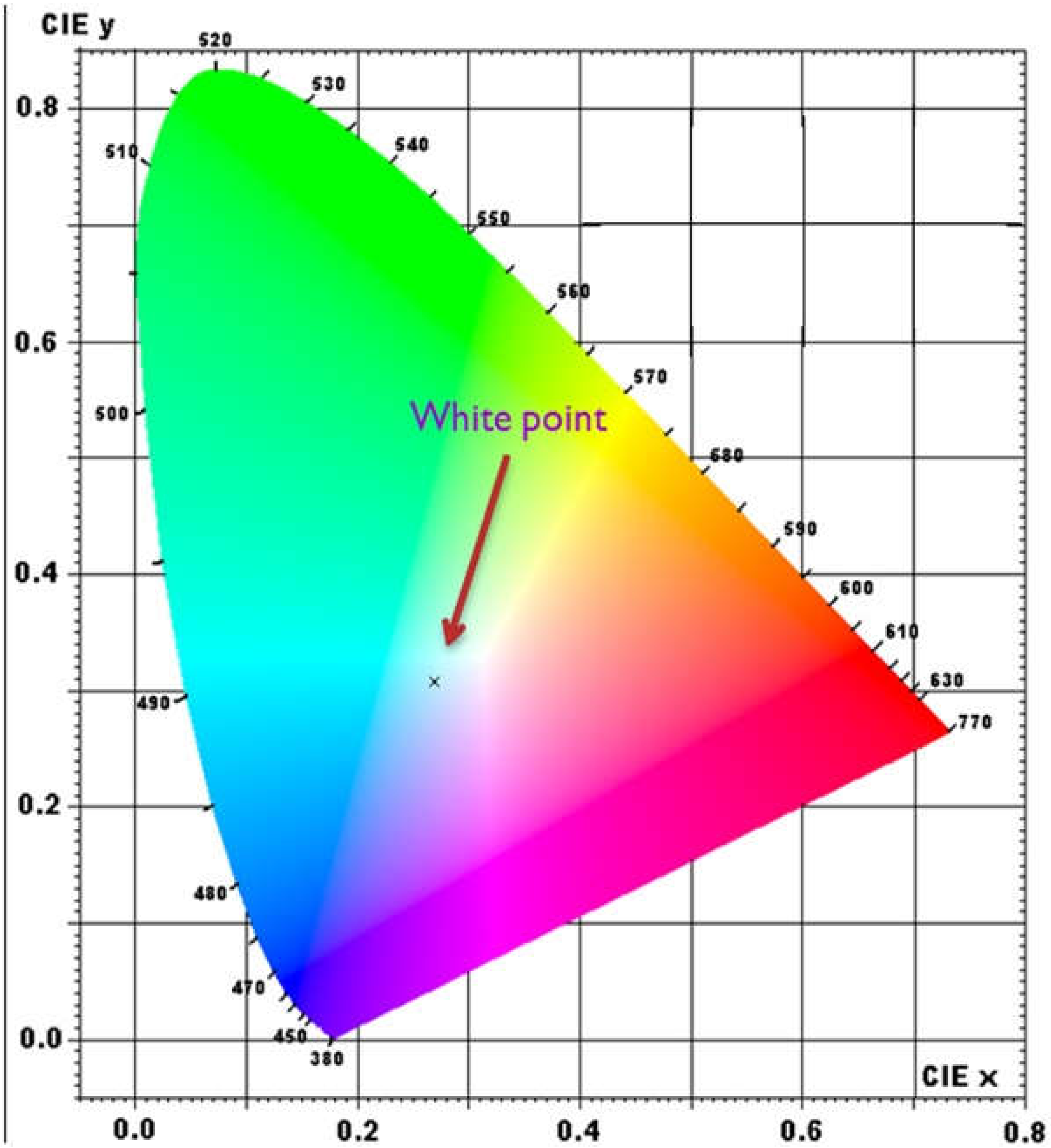